# Fast hierarchical inversion for borehole resistivity measurements in high-angle and horizontal wells using ADNN-AMLM

Yizhi Wu[a,b,c], Yiren Fan[a,b,c],*

[a] *School of Geosciences, China University of Petroleum, Qingdao 266580, China*

[b] *Laboratory for Marine Mineral Resources, Qingdao National Laboratory for Marine Science and Technology, Qingdao 266071, China*

[c] *CNPC Key Laboratory for Well Logging, China University of Petroleum, Qingdao 266580, China*

**Abstract** With the rapid development of deep learning, intelligent schemes have been gradually introduced to solve various inverse nonlinear problems. In this paper, we combine an efficient adaptive deep neural network (ADNN) framework with an adaptive modified Levenberg-Marquardt (AMLM) algorithm based on a three-layer inversion model to exact the formation resistivity and invasion depth from the measurements of array laterolog. The ADNN presented in this paper can achieve the 2D/3D fast forward modeling of array laterolog. The AMLM algorithm and a hierarchical inversion scheme are adopted to improve the anti-noise ability and convergence in complex logging environments, as well as achieve the fast and accurate reconstruction of longitudinal resistivity profiles in high-angle (HA)/horizontal (HZ) wells. Numerical simulations show that ADNN-based forward modeling only takes 0.021 s for each logging point, and the maximum relative error is less than 2%. The three-layer inversion model can eliminate the effect of the surrounding bed and improve the inversion accuracy in thinly layered formations. The AMLM inversion algorithm can effectively suppress the influence of noise, and takes only 10 steps to achieve convergence.

## 1 Introduction

In the drilling process of high-angle (HA)/horizontal (HZ) wells, the pressure difference between the drilling fluid and permeable layer will lead to the radial invasion of mud filtrate. Due to the differences in mud filtrate and the formation resistivities, the information obtained by dual lateral logging may not match the true resistivity of the formation (Liu et al.,1993; Mezzatesta et al.,1995; Zhao et al.,2019; Zhang et al.,2000), leading to inaccurate reservoir judgments. Therefore, Schlumberger developed a new high-resolution array lateral tool, HRLA (Itskovich et al.,1998), which provides six apparent resistivity measurements with different depths of investigation (DOIs) and is widely used in the evaluation of oil and gas reservoirs with HA/HZ wells. However, in thinly formations, the measured laterolog arrays are still seriously affected by the relative dipping angle between the well trajectory and the formation, drilling fluid invasion, the surrounding bed and other factors, thus, the true formation resistivity cannot be accurately estimated (Frenkel et al.,1995; Galli et al.,2002; Ni et al., 2017; Hu et al., 2019). Therefore, accurately extracting the real formation resistivity from the apparent resistivity information is important in evaluations of oil and gas reservoirs.

---

∗Corresponding author. School of Geosciences, China University of Petroleum, Qingdao 266580, China.
E-mail address: fanyiren@upc.edu.cn (Y. Fan).

Initially, to solve the resistivity distortion problem, a correction method was widely used by interpreters to eliminate the effects of boreholes, surrounding beds and invasion (Griffiths et al.,2000; Phelps et al.,2007; Maurer et al.,2009). However, this technique, which is based on single-factor superposition correction, is only applicable to thick layers in vertical wells (Abdel et al.,2011; Wu et al.,2019). Therefore, geophysical inversion technology that can extract the true resistivity of formation was introduced.

Inversion techniques are an excellent way to process HRLA data since they allow for the accurate and simultaneous consideration of both the formation resistivity distribution (radially and vertically) and the dipping angle. Due to the strong nonlinear relationship between the logging response and formation model, it is difficult to extract true values of formation parameters (invasion depth, invasion zone resistivity and undisturbed formation resistivity). Therefore, based on least squares theory, researchers applied the deterministic algorithm in laterolog array data processing and successfully reconstructed the formation resistivity profile (Hakvoort et al.,1998; Wang et al.,2009; Yao et al.,2010; Olabode et al.,2014; Hu et al.,2018; Wang et al.,2019). The objective of this algorithm is to obtain the global minimum through continuous iterations in the direction of gradient descent based on the objective function (Anderson et al.,2001; Yamashita et al.,2001). However, the deterministic algorithm has some disadvantages that cannot be ignored in 3D formation environments: (1) the 3D forward is time-consuming; (2) the calculation of the Jacobian matrix is memory-consuming;(3) the convergence and global optimization ability of the algorithm are strongly dependent on the initial values (Sergio et al.,2016; Li et al., 2019).

In addition, with the rapid development of intelligent algorithms, a new perspective for processing logging data has emerged. Compared with deterministic algorithms, intelligent optimization methods can not only eliminate the Jacobian matrix but also achieve global optimization based on an arbitrary initial value, which greatly improves the data processing efficiency (Ewa et al.,2015; Zhu et al.,2019; Feng et al.,2019; Wang et al.,2019; Li et al.,2020). Such algorithms include simulated annealing algorithms, differential evolution algorithms, particle swarm optimization algorithms and Bayesian algorithms. However, in all of these algorithms, forward modeling is still the key restricting factor for the inversion efficiency. Therefore, some researchers have introduced deep learning (DL) into geophysical inversion (Jin et al.,2019; Raissi et al.,2019; Bai et al.,2020). As the foundation of DL, deep neural networks (DNNs) technology has a strong learning ability and can be used as to establish surrogate models for forward simulation (Wu et al.,2019; Li et al.,2019; Kinjal et al.,2019). However, various inversion problems must still be solved, including how to obtain accurate logging databases, how to build efficient neural network frameworks, and how to ensure sufficient inversion precision.

In this paper, a new hierarchical joint inversion strategy is proposed for array laterolog measurements. Based on the 3D finite element method (3D-FEM), an efficient adaptive deep neural network (ADNN) framework is constructed, and the adaptive modified Levenberg-Marquardt (AMLM) algorithm with cubic convergence is introduced, which ensures the speed and accuracy of inversion for array lateral measurements. To simplify the workflow, a domain decomposition method is introduced to decompose the large logging data set into several subregions, which improves the inversion efficiency. The paper is organized as follows. In section 2, the principles of array laterolog, the 3D-FEM modeling method and the ADNN fast modeling method are introduced. In section 3, the AMLM algorithm, regional decomposition technology, uncertainty evaluation method and specific hierarchical inversion workflow are proposed. In section 4, synthetic examples are presented to verify the stability and efficiency of the inversion scheme. In the final section, we extend the inversion scheme to anisotropic formations.

## 2 Theory and forward method

In this section, we briefly review the operating mode of the array laterolog tool. Then the signal definition and forward modeling algorithm are presented.

*2.1 Physics of the HRLA*

The array laterolog tool provides measurements with different DOIs by using multifocusing configurations. Fig. 1 shows the basic framework of the HRLA introduced by Schlumberger in 1998, and it consists of a main electrode *A0*, six pairs of shielding electrodes *A1-A6* (*A1'-A6'*), and two pairs of monitoring electrodes *M1* and *M2* (*M1'* and *M2'*). These electrodes are symmetric with respect to the main current electrode *A0*. The current injected from the main electrode is focused by the adjacent shielding electrodes and then flows deeply into the formation. By changing the transceiver combinations, we can obtain 6 apparent resistivity curves with different DOIs, namely, *RLA0-RLA6*.

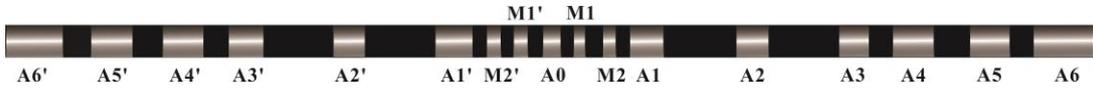

Fig. 1. Configuration of the HRLA

The apparent resistivities can be calculated by measuring the potential of the monitoring electrode. Taking the shallow detection mode *RLA0* as an example, the potential difference between two monitoring electrodes on the same side need to be measured, as shown in Eq. (1). In other detection modes, only the potential at a given monitoring electrode need to be measured, as shown in Eq. (1). The apparent resistivities are calculated as follows:

$$R_{RLA0} = K_{RLA0} \frac{\Delta U_{M1M2(RLA0)}}{I_{0(RLA0)}} \tag{1}$$

$$R_{RLAi} = K_i \frac{\Delta U_{M1(RLAi)}}{I_{0(RLAi)}} (i=1,2\cdots,5) \tag{2}$$

where, *i* represents the detection mode; $R_{RLA0}$ and $R_{RLAi}$ are the apparent resistivities under different detection modes; $K_{RLA0}$ and $K_i$ are the electrode system coefficients of the HRLA under different detection modes; $\Delta U_{M1M2}$ is the potential difference between monitoring electrodes (*M1* and *M2*); $U_{M1}$ is the potential of *M*1; and $I_{0(RAL0)}$ and $I_{0(RAL0)}$ are the emission current intensities of the main electrode $A_0$.

The key to determining the apparent resistivity of the HRLA is to obtain the potential of the monitoring electrode *M*1 at any position in the 3D domain, thus, the potential distribution function of the whole space can be calculated.

*2.2 Numerical Simulation Method*

Since the HRLA is powered by a low-frequency alternating current, the logging response can be reduced to a direct-current field calculation. In the cylindrical coordinate system ($r$, $\varphi$, $z$), the differential equation can be expressed as (Zhang, 1996):

$$\frac{\partial}{\partial r}\left(\frac{r}{\rho}\frac{\partial U}{\partial r}\right)+\frac{\partial}{\partial \varphi}\left(\frac{1}{Rr}\frac{\partial U}{\partial \varphi}\right)+\frac{\partial}{\partial z}\left(\frac{1}{\rho}\frac{\partial U}{\partial z}\right)=0 \tag{3}$$

where $U$ is the measured potential and $\rho$ is the resistivity of the formation.

It should be noted that the current and potential are continuous at the interface of the subspace with different resistivities, which satisfies the following boundary conditions:

$$\begin{cases} U_- = U_+ \\ \left[\dfrac{1}{\rho}\dfrac{\partial U}{\partial \vec{n}}\right]_- = \left[\dfrac{1}{\rho}\dfrac{\partial U}{\partial \vec{n}}\right]_+ \end{cases} \tag{4}$$

where, '−' and '+' represents the possible directions at the interface and $\vec{n}$ is the unit normal vector at the interface.

Furthermore, Dirichlet and Neumann boundary conditions must be satisfied at the outer boundary and the surface of the insulated electrode:

$$\begin{cases} U\big|_{\Gamma_1} = 0 \\ \dfrac{\partial U}{\partial \vec{n}}\bigg|_{\Gamma_2} = 0 \end{cases} \tag{5}$$

where, $\Gamma_1$ and $\Gamma_2$ represent the outer boundary and the surface of the insulated electrode, respectively.

On the surfaces of the main electrode and shielded electrodes, the equipotential boundary conditions should be satisfied:

$$-\iint_{\Gamma_3} \frac{1}{\rho}\frac{\partial U_{A_i}}{\partial \vec{n}} dS = I_i \tag{6}$$

where, $U_{A_i}$ is the potential of electrode $A_i$, $\rho$ is the resistivity of the region where the electrode $A_i$ is located, $\Gamma_3$ is the surface of the main or a shielded electrode, $\vec{n}$ is the unit normal vector at the interface of $A_i$ and $I_i$ is the emission current intensity.

To obtain Eq. (3), the partial differential equation-based problem is transformed into an extreme function-based problem:

$$\phi(U) = \frac{1}{2}\iiint_\Lambda \frac{1}{\rho}[(\frac{\partial U}{\partial r})^2 + \frac{1}{r^2}(\frac{\partial U}{\partial \varphi})^2 + (\frac{\partial U}{\partial z})^2]d\Lambda - \sum_{i=0}^{6}I_iU_i \qquad (7)$$

where, $\Lambda$ is the solution area, $I_i$ and $U_i$ respectively are the current and potential of each electrode.

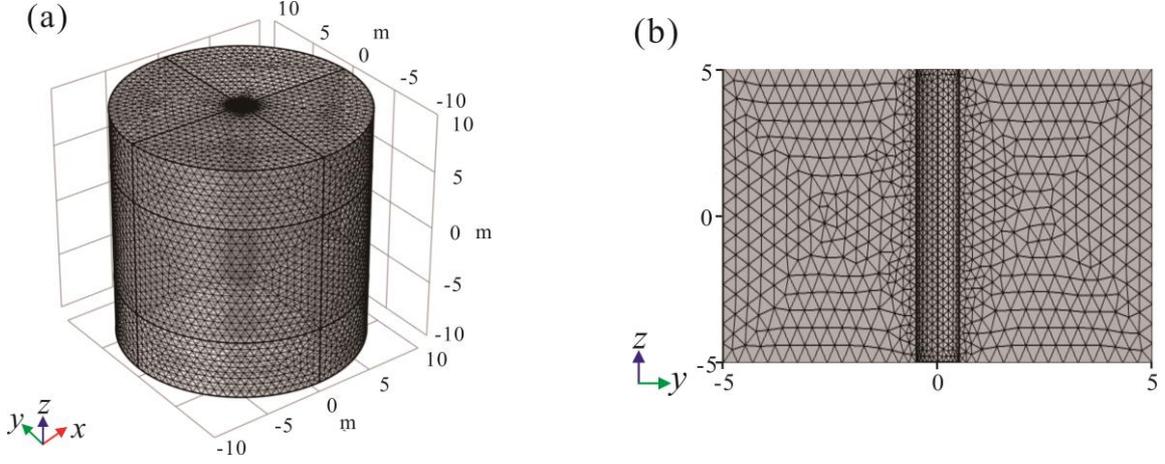

Fig. 2. FEM mesh: (a) Mesh of a 3D cylinder model; (b) Mesh structure at $x=0$ in the 3D cylinder model

To obtain the minimum value of the above function, the finite element method (FEM) is adopted in this paper to discretize the solution area. In the direction of $r$, the equal-spacing division method is used inside the borehole and nonequal-spacing is used in the formation. In the $r$-$\theta$ plane, equal-spacing is used for division and the $z$ direction is divided with nonuniform spacing. In this way, the solution region is divided into a certain number of hexahedrons. However, in the three-dimensional finite element method (3D-FEM), tetrahedrons are widely used because of their high calculation accuracy and adaptability, as shown in Fig. 2. Therefore, we further divide these hexahedra into tetrahedra. Thus, the continuous differential equations and boundary conditions with infinite degrees of freedom are transformed into equations containing finite node variables. Finally, the front-line solution method is adopted to solve the linear equations with large sparse matrix, and the potential values of each node can be obtained:

$$K\Phi = F \qquad (8)$$

where, $K = \sum_{i=1}^{m} K^{e_i}$, $m$ is the number of elements; $\Phi = [U_1,\cdots,U_N]^T$, $F = [I_1,\cdots,I_N]^T$ and $N$ is the number of nodes.

Since the number of shielded and returned electrodes can vary when the instrument is in different detection modes, the principle of electric field superposition is used to calculate the real electric field. The objective of the method is to generate 7 subfields by assuming that every electrode in the HRLA independently emits current. It should be noted that the current injected by main electrode A0 is assumed to be 1A, and the current intensities injected by the other 6 electrodes are $I_{A1}$~$I_{A6}$, respectively. Finally, seven subfields are superposed into real electric fields:

$$U = U_0 + I_{A1}U_1 + I_{A2}U_2 + I_{A3}U_3 + I_{A4}U_4 + I_{A5}U_5 + I_{A6}U_6 \tag{9}$$

where, $U_0$~$U_6$ are the distribution functions of the seven subfields.

To obtain the six weighted coefficients in Eq. (9), six equations generated based on the six detection modes of the HRLA should be satisfied. Taking Mode 6 (*RLA5*) as an example, the two monitoring electrodes on the same side should have the same electric potential, the shielded or backflow electrodes should have the same electric potential, and the instrument current should be constant. The corresponding formula set is expressed as follows:

$$\begin{cases} U_{M1} = U_{M2} \\ U_{A1} = U_{A2} = U_{A3} = U_{A4} = U_{A5} \\ I_{A0} + I_{A1} + I_{A2} + I_{A3} + I_{A4} + I_{A5} + I_{A6} = 0 \end{cases} \tag{10}$$

where, $U_{M1}$ and $U_{M2}$ are the potential of monitoring electrodes *M*1 and *M*2, $U_{A1}$~$U_{A5}$ are the potential of shielding electrodes *A1*~*A5*.

In this paper, to verify the accuracy of the 3D-FEM algorithm, we compare the results simulated by 3D-FEM with those of the physical simulation software COMSOL. The random formation model is given as follows. The borehole diameter is 8 inches, and the borehole is filled with drilling mud with a resistivity of 0.1 Ω.m. The horizontal and vertical resistivities of the middle layer are 20 Ω. m and 80 Ω. m, respectively. In addition, the dipping angle is 80°. The resistivity of the surrounding layer on both sides is 2 Ω.m. The simulation results are shown in Fig. 3a. The scatter points illustrate the COMSOL simulation results and the solid line is the 3D-FEM simulation results. Fig. 3b shows the mean relative error (MRE) at each test point. The maximum error is less than 0.2%, which reflects the high accuracy of 3D-FEM.

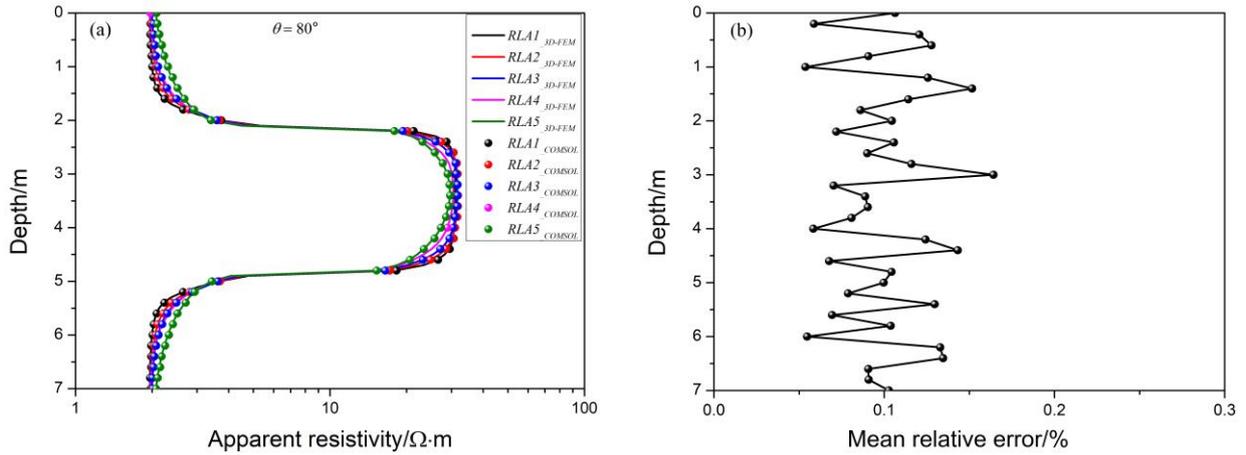

Fig. 3. (a)Simulated results of 3D-FEM and COMSOL and (b)MRE between them

*2.3 Rapid forward modeling method*

*2.3.1 Deep neural network theory*

Traditionally, inversion is time-consuming because hundreds or thousands of 3D-FEM modeling function invocations are involved. In this paper, we propose a rapid forward-modeling approach based on the ADNN framework.

Due to their strong generalization characteristics and learning capabilities, deep neural networks (DNNs) are widely used in DL. Specifically, multilayer neural networks, mature DL frameworks, have been widely applied in natural language processing, image target detection, and data prediction (LeCun et al.,2015). A classic multilayer neural network is composed of three layers, as shown in Fig. 4, including an input layer, a hidden layer and an output layer. The neurons in each layer are connected by weights, biases and activation functions. Generally, the three-layer neural network (TL-NN) framework can approximate a simple continuous function. However, for the forward modeling of array laterolog, which contains multiple inputs and outputs, the learning capability of a TL-NN is too weak, and a TL-NN cannot be used as a surrogate model. Therefore, in this paper, an ADNN is constructed to predict the responses of the HRLA in HA/HZ wells. The specific process is as follows: (1) label the formation model, and create a database; (2) train the database, and verify the accuracy of the ADNN; (3) predict the responses of the HRLA.

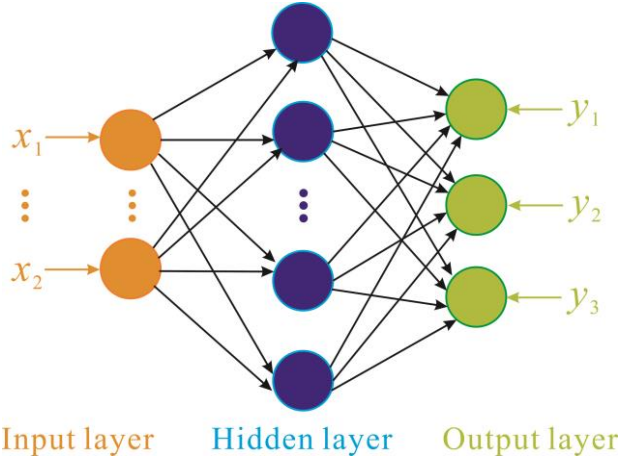 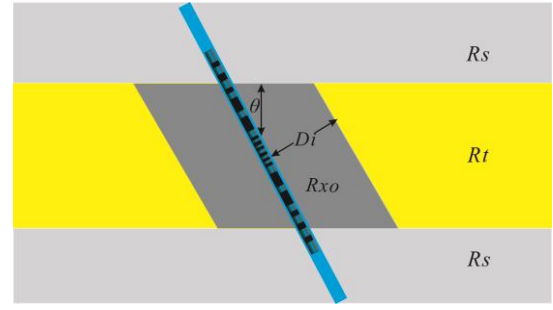

Fig. 4. Three-layer neural network framework    Fig. 5. Three-layer training or inverted model

*2.3.2 Rapid forward modeling*

When drilling through a thinly layered bed in HA/HZ wells, the effect of the surrounding bed on the array laterolog responses cannot be ignored. Therefore, a three-layer model, as shown in Fig. 5, is considered in DL. There are six parameters in the model, including the dipping angle ($\theta$), the resistivities of the surrounding and target layers ($Rs$ and $Rt$), and the invasion depth ($Di$), resistivity ($Rxo$), and thickness of the target layer ($H$). To decrease the number of neurons in the input layer, the training model is divided into 10 submodels with different dipping angles. The range of dipping angles is from 0° to 90°, and the interval is 10°. Furthermore, a sand-shale sequence is mainly considered in this paper, and it is assumed that the resistivity of the surrounding bed is low. Therefore, the ranges of parameters used in every submodel are expressed in Table 1. In this manner, the database for each submodel is established based on 3D-FEM. It should be noted that we randomly selected 95% of the data for training and the other data for validation of the network.

Table 1 The ranges of parameters used in the trained model

| Parameter | $Di$ (m) | $H$ (m) | $Rt$ (Ω.m) | $Rs$ (Ω.m) | $Rxo$ (Ω.m) |
|---|---|---|---|---|---|
| Range | 0 ~ 2 | 0~5 | 0.1~200 | 0.1~50 | 0.1 ~30 |

In the ADNN, the loss function $L(x)$ is defined based on maximum likelihood estimation, as shown in Eq. (11), and the sigmoid function, as shown in Eq. (12), is set as the activation function.

$$L(w,b,x,p) = -\frac{1}{n}\sum_{i=1}^{n}\left(y*log(g(w(p),b(p),x)) + (1-y)*log(1-g(w(p),b(p),x))\right) \quad (11)$$

$$\sigma(z) = \frac{1}{1+exp(-z)} \quad (12)$$

where, $g(w,b,x) = \sigma(w*x+b)$, $w$ and $b$ are the weights and biases, respectively; $p$ is the number of iterations; $n$ is the number of samples; $x$ is the vector of the input layer (here, $x=(Rs,Di,Rxo,Rt,H)^T$); and $y$ is the vector of the output layer (here, $y = log(RLA1,RLA2,RLA3,RLA4,RLA5)^T$).

There are two learning mechanisms in the ADNN, including a feedforward mechanism (FF) and a back-propagation mechanism (BP). The loss function can be calculated through the FF mechanism. Then, the weights and biases of the ADNN will be adjusted through BP based on the chain rule until the loss function reaches a manually selected accuracy threshold, as shown in Eq. (13).

$$\begin{cases} w^k(p+1) = w^k(p) - \alpha \dfrac{\partial L(w^k(p), b^k(p))}{\partial w^k(p)} \\ b^k(p+1) = b^k(p) - \alpha \dfrac{\partial L(w^k(p), b^k(p))}{\partial b^k(p)} \end{cases} \quad (13)$$

where, $k$ is the number of hidden layers and $\alpha$ is the learning rate; generally, $\alpha<1$.

In addition, to accelerate the training of the ADNN, the momentum term is introduced in each iteration, as follows:

$$\begin{cases} w^k(p+1) = w^k(p) - \alpha \dfrac{\partial L(w^k(p), b^k(p))}{\partial w^k(p)} - \eta\alpha \dfrac{\partial L(w^k(p-1), b^k(p-1))}{\partial w^k(p-1)} \\ b^k(p+1) = b^k(p) - \alpha \dfrac{\partial L(w^k(p), b^k(p))}{\partial b^k(p)} - \eta\alpha \dfrac{\partial L(w^k(p-1), b^k(p-1))}{\partial b^k(p-1)} \end{cases} \quad (14)$$

where $\eta$ is the momentum coefficient in the interval (0, 1).

Then, an adaptive learning rate is introduced to improve the training accuracy and speed:

$$\alpha(p+1) = \zeta\alpha(p) \quad (15)$$

$$\begin{cases} \zeta \geq 1 & \text{if } L(w,b,x,p+1) < L(w,b,x,p) \\ 0 < \zeta < 1 & \text{otherwise} \end{cases} \quad (16)$$

where, $\zeta$ is the adaptive coefficient and $\alpha(p)$ is the learning rate for the $p^{th}$ iteration.

Finally, a genetic algorithm (GA) is adopted to search the suitable weights and biases through BP, which ensures the accuracy of the ADNN. A detail introduction to GA was provided by Byungwhan (2005).

Now, it is convenient to present the complete algorithm as follows.

**Algorithm of the ADNN**

**Input:** $Rs$, $Di$, $Rxo$, $Rt$ and $H$. Maximum iterations $Ite = 10000$; Initial learning rate $\alpha_0 = 0.1$; $\eta = 0.8$; Terminates iteration threshold $Goal = 1E-6$;

**Step 1.** Normalize the input data

**Step 2.** Set $p = 1$, and initialize the weights and biases using the GA

**Step 3.** Compute $L(w, b, x, p)$. If $L<Goal$, Stop; otherwise, go to Step 4.

**Step 4.** Calculate $w^k(p+1)$ and $b^k(p+1)$ based on (14)

**Step 5.** Calculate $L(w, b, x, p+1)$ based on (11)

**Step 6.** Choose $\zeta$ as follows:

$$\begin{cases} \zeta \geq 1 & \text{if } L(w,b,x,p+1) < L(w,b,x,p) \\ 0 < \zeta < 1 & \text{otherwise} \end{cases}$$

**Step7.** Set $\alpha(p+1) = \alpha(p)$

**Step8.** Set $p = p+1$. If $k<=Ite$, go to Step 3; otherwise, stop.

**Step9.** Output and save the ADNN architecture

*2.3.3 Precision validation of the ADNN*

In this section, we test the precision of the trained ADNN. The borehole diameter of the testing model is 8 inches filled with drilling mud with a resistivity of 0.1 Ω.m, and the other parameters are expressed in Tabel 2. The measurement point is located at the middle of model.

Table 2 Parameters of the testing model

|  | $Di$ (m) | $\theta$ (degree (°)) | $H$ (m) | $Rt$ (Ω.m) | $Rs$ (Ω.m) | $Rxo$ (Ω.m) |
|---|---|---|---|---|---|---|
| Model1 | 0 ~ 1.5 | 0 | 3 | 50 | 2 | 3 |
| Model2 | 0.5 | 0 ~ 90 | 3 | 20 | 2 | 3 |
| Model3 | 0.5 | 0 | 0~5 | 50 | 2 | 3 |
| Model4 | 0.5 | 0 | 3 | 3 ~ 150 | 2 | 3 |

The effects of $Di$, $\theta$, $H$ and $Rt$ on the prediction accuracy of the ADNN are analyzed in Fig. 6(a)-(d). The scatter points are the predicted results of the ADNN, and the solid line is the 3D-FEM simulation results. The predicted results of the ADNN agree well with the simulation results of 3D-FEM, and the maximum relative errors in Fig.6 are 1.93%, 0.02%, 0.88% and 0.39%, respectively. In addition, it takes approximately 0.021 s to compute values for one logging point with the ADNN; this computational time is nearly 50 times faster than 3D-FEM algorithm.

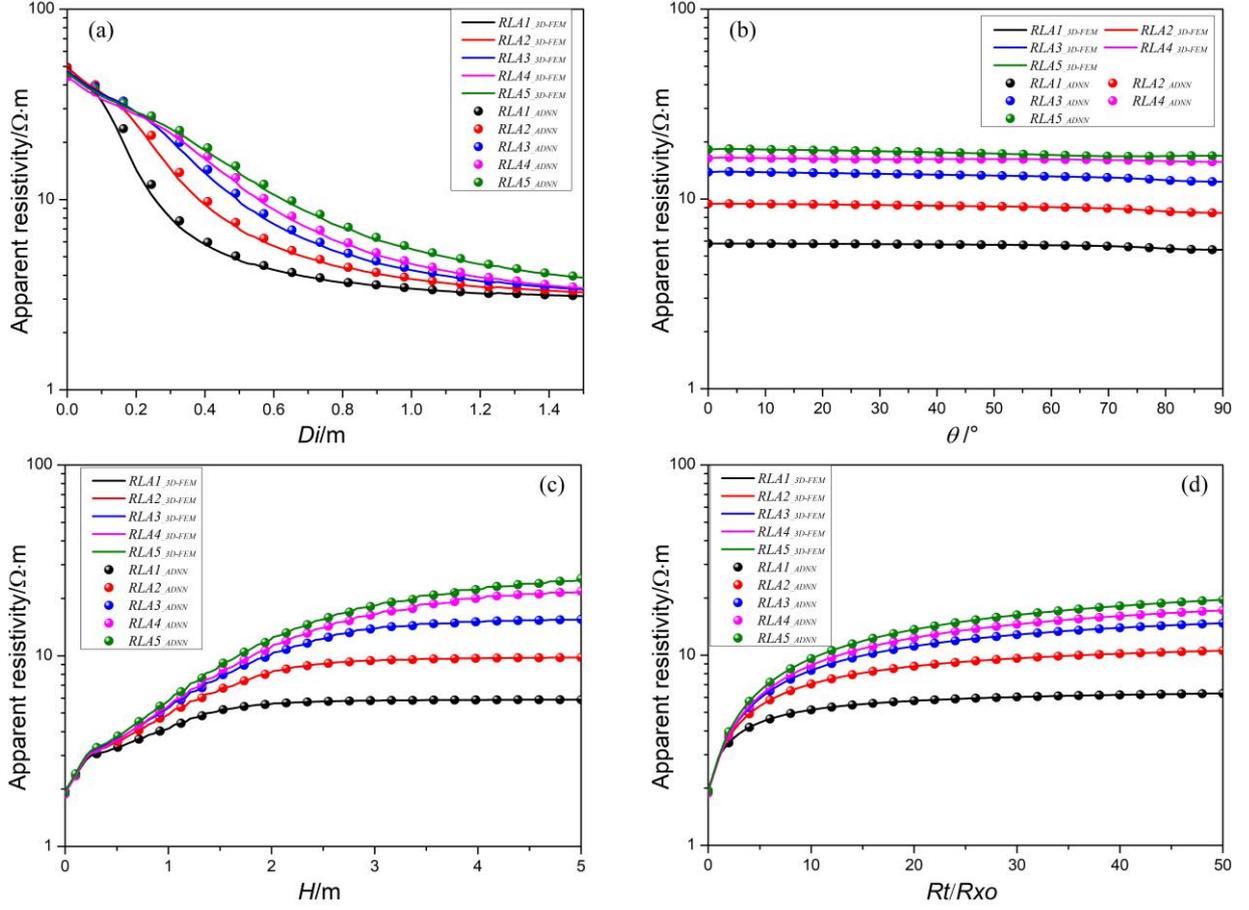

Fig. 6. Comparison of the simulated results of the ADNN and 3D-FEM in various formations with different (a) *Di*, (b) *θ*, (c) *H* and (d) *Rt/Rxo* values

## 3 Inversion method

*3.1 Adaptive modified Levenberg-Marquardt algorithm*

The traditional Levenberg-Marquardt (L-M) algorithm can improve the global convergence of inversion because it combines the Gauss-Newton approach with the steepest descent method. However, the observed measurements of array laterolog are complicated by the effects of noise, which can limit the global convergence and the efficiency of inversion. Therefore, the adaptive modified Levenberg-Marquardt method (AMLM) with cubic convergence is proposed to optimize anti-noise performance and convergence. The cost function in inversion is as follows:

$$C(x) = \frac{1}{2}\sum_{i=1}^{m}[f_i(x)]^2 = \frac{1}{2}\left\{\|e(x)\|^2 + \|\lambda(x-x_p)\|^2\right\} \qquad (17)$$

where, *x* is the inverted vector of the $N^{th}$ layer model, $x = (Di_1, Rxoh_1, Rth_1, \lambda_1, \cdots, Di_n, Rxoh_n, Rth_n, \lambda_n)^T$, $T$ is the transpose of matrix; $\|e(x)\|^2$ is the *L2* norm of the difference between the measured and forward modeling values; $x_p$ is a reference vector containing the parameters for step *p*-1; and $\lambda$ is a regularization parameter. The final term in the cost function is used to suppress measurement noise and eliminate pathological effects in inversion. The purpose of inversion is to obtain an *x\** value that minimizes the cost function; i.e., *x\** = argmin{*C(x)*}.

At every iteration, the trial step $d_k$ is executed as in the L-M method:

$$d_k = -J^T(x^k)C(x^k)\left[J^T(x^k)J(x^k) + \lambda_k I\right]^{-1} \tag{18}$$

where, *k* is the iteration number, *I* is the identity matrix and $J(x^k)$ is the Jacobian at $x^k$.

If *J(x)* is Lipschitz continuous and nonsingular at the solution, then the convergence of the L-M method is quadratic. However, it is difficult to satisfy the inherent nonsingular characteristics in the inversion of array laterolog with HA/HZ wells. Therefore, Fan (2012) introduced the modified L-M method(MLM) with cubic convergence. The specific modified approach involves executing a pseudotrial step $d_k$ at each iteration:

$$d_k = -J^T(x^k)C(y^k)\left[J^T(x^k)J(x^k) + \lambda_k I\right]^{-1} \tag{19}$$

where, $y^k$ is the solution of $(k+1)^{th}$ step in LM algorithm, $y^k = x_k + d^k$ and $\lambda_k$ is the regularization parameter.

In addition, an adaptive regularization parameter selection method was proposed by Keyvan Amini (2015) to optimize global convergence:

$$\delta_k = \begin{cases} \dfrac{1}{\|C(x_k)\|} & \text{if } \|C(x_k)\| \geq 1, \\ 1 + \dfrac{1}{k} & \text{otherwise} \end{cases} \tag{20}$$

$$\lambda_k = \omega_k \|C(x_k)\|^{\delta^k} \tag{21}$$

where, $\omega_k$ is the adaptive coefficient of the regularization parameter, which is calculated as follows：

$$\omega_k = \begin{cases} 4\omega_{k-1} & \text{if } r_k < p_1, \\ \omega_{k-1} & \text{if } r_k \in [p_1, p_2], \\ \max\{\omega_{k-1}/4, m\} & \text{otherwise} \end{cases} \tag{22}$$

where, *m* is a positive constant and $r_k$ is the gain coefficient, which is defined as follows：

$$r_k = \frac{\|C(x^k)\| - \|C(x^k + d_k + \alpha_k d_k)\|}{\|C(x^k)\| - \|C(x^k) + J(x^k)d_k\| + \|C(y^k)\| - \|C(y^k) + \alpha_k J(x^k)d_k\|} \tag{23}$$

Therefore, the final trial step at the $k^{th}$ iteration $s_k$ is:

$$\begin{cases} s_k = d_k + \alpha_k \cdot d_k \\ \alpha_k = min\left\{1 + \dfrac{\lambda_k d_k^T d_k}{d_k^T J^T(x_k) J(x_k) d_k}, \alpha\right\} \end{cases} \tag{24}$$

where, $\alpha$ is the constraint factor, generally, $\alpha > 1$.

To verify the robustness of the AMLM method, an invaded formation model with $Di = 0.5$ m is given in this paper. The resistivities of the invaded and uninvaded zone are 2 $\Omega.m$ and 20 $\Omega.m$, respectively. The inversion parameter vector is $x = (Di, Rxo, Rth)^T$, and two initial value vectors are $x_1 = (0.8, 50, 100)^T$ and $x_2 = (0.3, 10, 50)^T$. The cost function varies with the number of iterations and is shown in Fig. 7.

Notably, Fig. 7a shows that the AMLM algorithm will converge faster than the L-M method when the initial value is close to the true value. Additionally, Fig. 7b shows that the AMLM algorithm improves the global convergence when the initial value is 25 times larger than the true value.

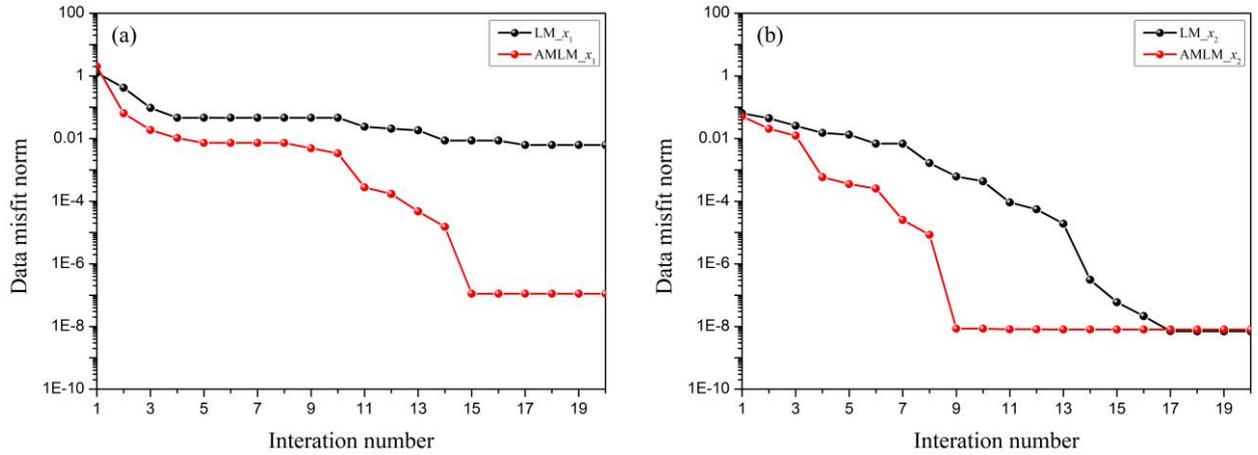

Figure. 7. Evolution of the data fitting (misfit) norm using the (a) AMLM and (b) LM methods

*3.2 Hierarchical joint inversion scheme and inversion window design*

In this paper, a hierarchical joint inversion scheme is proposed to simplify and accelerate the inversion of measured laterolog arrays in HA/HZ wells. In the first stage of inversion, 3D-FEM modeling is replaced with ADNN modeling to obtain the first inverted results and reduce time consumption based on the AMLM method. Then, in the second stage of inversion, as a strict forward algorithm, 3D-FEM is used to guarantee the precision of the inverted results. Because the initial values in the second stage are close to the true initial values, this step is also not time consuming.

In this work, we adopt the domain decomposition method to divide the entire logging data set corresponding to long depth intervals that include a significant number of formation layers into many windows, as shown in Fig. 8. It should be noted that the three-layer model is included in each window. The thickness of the middle layer can extend to the boundaries, which can be determined by the inflection points of measurements, gamma rays or spontaneous potentials. In addition, the thicknesses of the upper and lower layers are set to semi-infinite in the inversion process. To improve the inversion precision, the inverted results for the middle and lower layers in the $i^{th}$ window are set as the initial values of the upper and middle models in the $(i+1)^{th}$ window. The initial values of the lower model in the $(N+1)^{th}$ window are artificially assigned. In particular, the inversion process can be repeated two or more times until the stopping iteration conditions are satisfied. After the $j^{th}$ processing step for a whole data set, the inverted results will be regarded as the initial values of the $(j+1)^{th}$ processing step to improve the accuracy of the results. The inversion workflow is shown in Fig. 9.

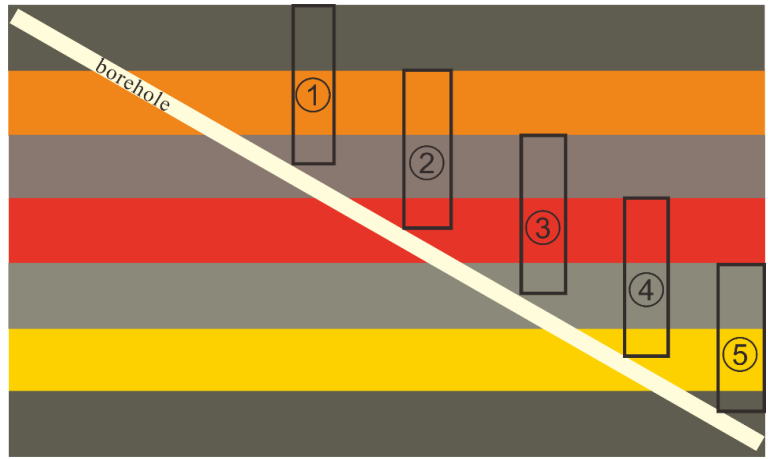

Figure. 8. Domain decomposition scheme

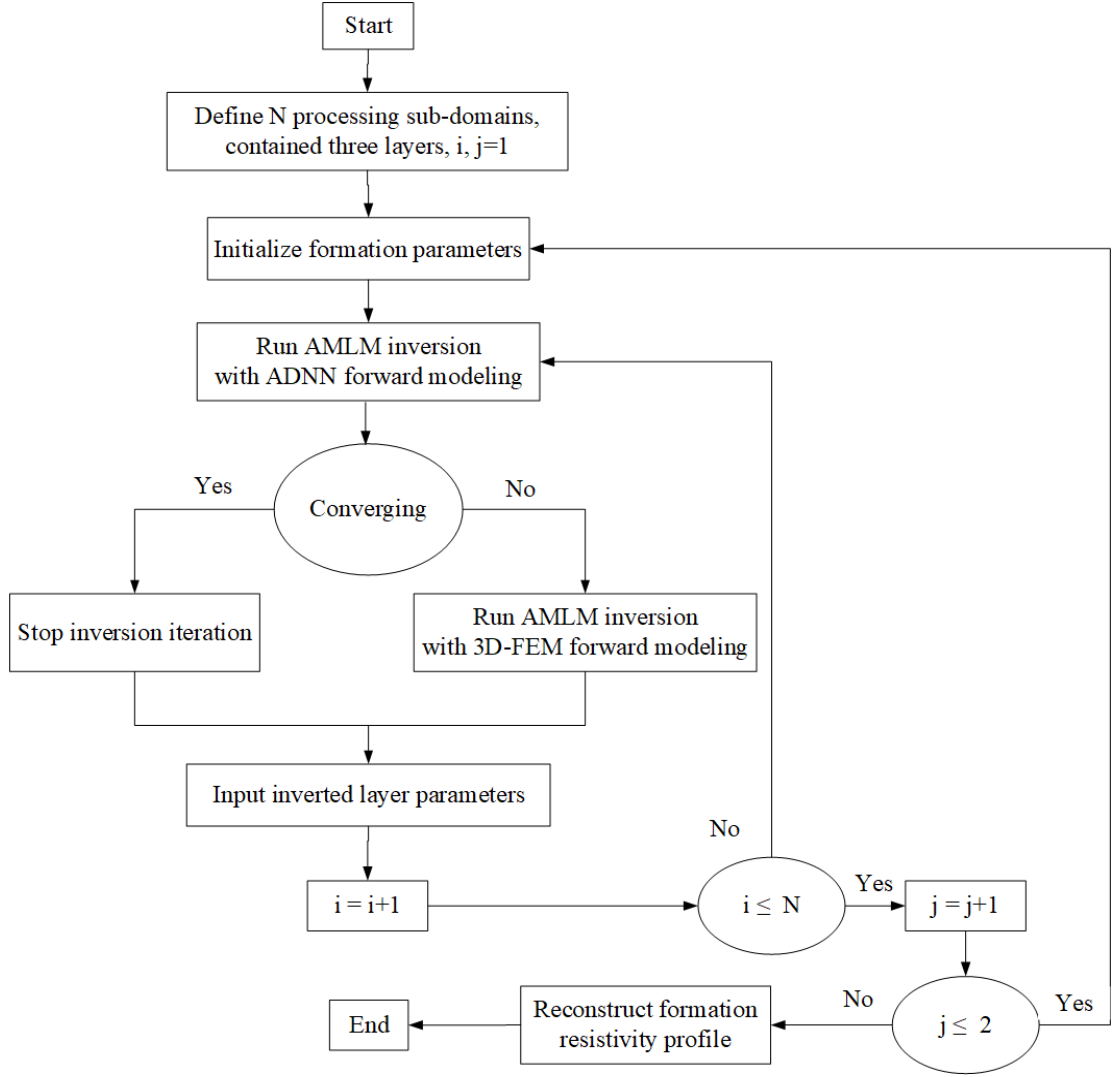

Figure. 9. Workflow of the hierarchical joint inversion scheme

*3.3 Uncertainty evaluation method*

In this paper, the covariance matrix is adopted to evaluate the uncertainty of the inversion results, and the formula is as follows:

$$\Theta = E\left\{\left[x^* - E(x^*)\right]\left[x^* - E(x^*)\right]^T\right\} \tag{25}$$

where, $x^*$ is the final solution of inversion and $E$ denotes the expectation value.

When the noise satisfies the Gaussian distribution, $\Theta$ can be approximated by the Fisher information matrix:

$$\Theta \approx \left[\Theta_s^{-1} + \xi J^T(x^*)\Theta_d^{-1}J(x^*)\right]^{-1} \tag{26}$$

where, $\xi = 1/\theta_d^{-2}\lambda^*$; $J(x^*)$ is the Jacobi matrix corresponding to the solution $x^*$; $s$ and $d$ are the numbers of inverted parameters and data points, respectively; $\Theta_s = I_s$; $\Theta_d = \theta_d^2 I_d$; $\theta_d$ is the standard deviation of the data points; $I_d$ and $I_s$ are identity matrices $I_s \in R^{s \times s}$ and $I_d \in R^{d \times d}$, respectively.

The Jacobi matrix can be decomposed by the singular value decomposition method as $J = USV^T$. Additionally, Eq. (26) can be simplified as follows:

$$\Theta \approx V\left(I_m + \frac{1}{\lambda^*}S^2\right)V^T \qquad (27)$$

where, $I_m$ is an identity matrix and $\lambda^*$ is the regularization parameter.

Finally, the uncertainty of parameters in the $i^{th}$ layer can be evaluated based on the square root of the diagonal terms of the estimator's covariance matrix:

$$\rho_i = \sqrt{\Theta_{ii}} \qquad (28)$$

## 4 Synthetic inversion experiments

In this section, 2D synthetic data generated by HRLA are processed to verify the effectiveness of the new proposed inversion scheme, and the detailed implementations are given. Pseudo random noise of different intensities is added to the synthetic data.

*4.1 Synthetic example 1: inversion for measurements without noise*

The well-known Oklahoma (OK) benchmark models with different relative dipping angles ($\theta$), illustrated in Fig. 10a-b, are represented to show the effectiveness of the inversion scheme. The model has 17 layers in a depth segment of 30 m. The formation includes multilayer invasion and consists of alternating conductive and resistive layers with thicknesses (*H*) from 0.3 m to 4.0 m. The borehole diameter is 8 inches filled with drilling mud with a resistivity of 0.1 Ω.m. The invasion depth *Di* varies from 0 m to 1.3 m. The resistivity of invaded zones *Rxo* varies from 0.2 Ω.m to 10 Ω.m. The resistivity of uninvaded zones *Rt* varies from 0.2 Ω.m to 100 Ω.m. Fig. 10c-d show the corresponding measurements simulated with the 3D-FEM code, and the dipping angles are 30º and 80º.

From the responses shown in Fig. 10c-d, the apparent resistivities are distorted because of the effects of the surrounding thinly layered beds, e.g., the $2^{nd}$ bed. In addition, due to the effect of mud invasion, the apparent resistivities are more than five times lower than the true formation resistivity, even in thick beds, e.g., the $6^{th}$, $8^{th}$ and $15^{th}$ beds. This characteristic becomes increasingly obvious as the dipping angle increases. Therefore, it is necessary to apply inversion to measurements to reconstruct the true resistivity profile of the formation.

In this case, we assume that the locations of formation boundaries are determined from other logging information and only invert the invaded-zone resistivity $R_{xo}$, uninvaded-zone resistivity $R_t$, and invasion radius $D_i$. The inversion process is executed two times for a 30 m segment with 0.1 m sampling interval. The time consumption is 87 s on a single i7-9700 processor with 8 GB of RAM. Fig. 11 shows the inverted results based on the inversion scheme proposed in this paper.

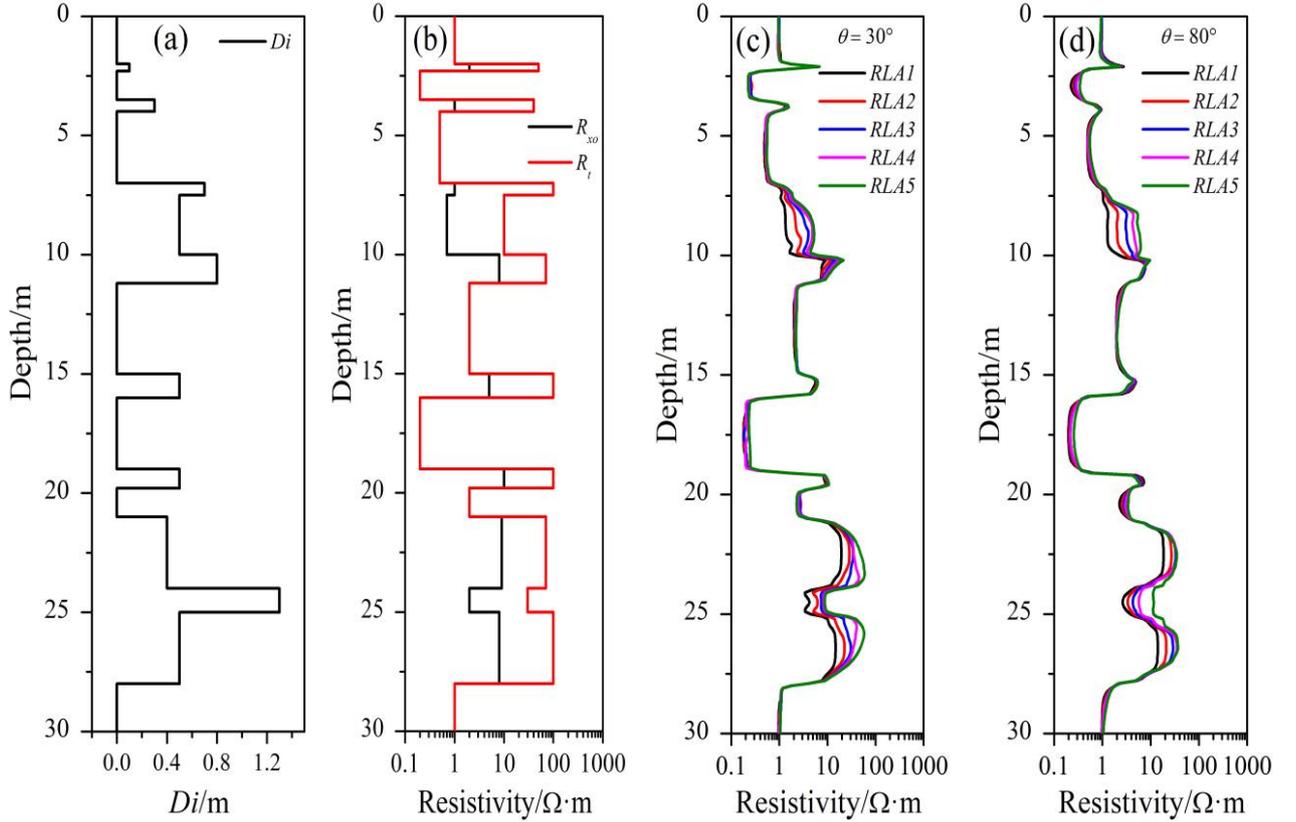

Fig. 10. OK model with different (a)Di, (b)$R_{xo}$ and $R_t$ and array laterolog responses for dipping angles of (c) $30^0$ and (d) $80^0$

Fig. 11 shows that the inverted $R_t$ values are more accurate than the inverted $R_{xo}$ values. The mean relative errors for $R_t$ and $R_{xo}$ are 4.11% and 6.68%, respectively. With increasing dipping angle, the logging response is also affected by the surrounding beds, which increases the relative error. In general, the mean relative errors of the inverted results at 30° and 80° are 3.46% and 4.66%, respectively. Overall, the inverted results agree well with the original formation values. It is worth noting that the inversion model adopted in this paper includes invasion in every layer. Due to the existence of multiple solutions, the inverted $D_i$ in an uninvaded bed is not zero, but the inverted $R_{xo}$ and $R_t$ values are very close to zero, e.g., as observed for the 9[th] bed.

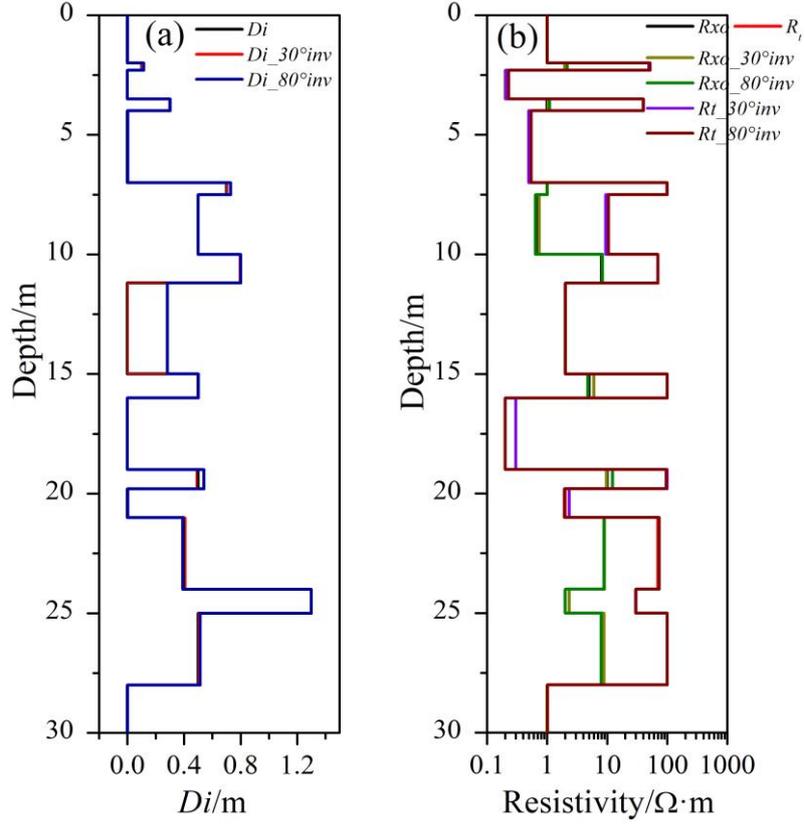

Figure. 11. Model and inverted (a) *Di* and (b) *Rxo* and *Rt* at dipping angles of 30º and 80º

*4.2 Synthetic example 2: inversion for measurements with pseudorandom noise*

To restore the complex downhole environment and investigate the effects of noise on the inversion scheme, we assume that the measurements are contaminated by pseudorandom noise and that the inversion parameters include *Di*, *Rxo* and *Rt*.

Both the formation boundaries and synthetic data are contaminated with noise as follows:

$$z_j = z_j + \gamma\varepsilon, \quad j = 1,\cdots,16$$
$$Ra_{mn}^{obs} = Ra_{mn}^{obs}(1+\gamma\varepsilon), \quad m = 1,\cdots,M; n = 1,\cdots,N \tag{28}$$

where, $z_j$ is the boundary position and $\gamma$ specifies the intensity of pseudorandom noise. $\varepsilon$ is random noise realized with a uniform distribution in the interval [1, 1].

Based on an *OK* benchmark model with a 30° dipping angle and a 3% pseudorandom noise added to the boundary positions, Fig. 12 shows the measurements for different intensities. From the left panel to the right panel in Fig. 12, the intensities are 5%, 10% and 25%. The inverted results are shown in Fig. 13. It is obvious that the stronger the noise is, the larger the relative error. Specifically, we observe that *Rxo_inv*, the inverted resistivity of the invaded zones, is generally more accurate than *Rt_inv*, the inverted resistivity of the uninvaded zones. When the noisy intensity is 25%, the mean relative errors from top to bottom in the *OK* model for *Rxo* and *Rt* are 7.72% and 9.11%, respectively. Furthermore, the influence of noise on *Rt* in thinly layered beds cannot be completely eliminated, e.g., as observed for the 15$^{th}$ bed. The relative error for *Rt* is 33.64%.

Generally, in this case, although the boundary positions and measurements are contaminated with noise, the inverted results based on the hierarchical joint inversion scheme still agree well with the values in the original formation model, which verifies the robustness of the scheme proposed in this paper.

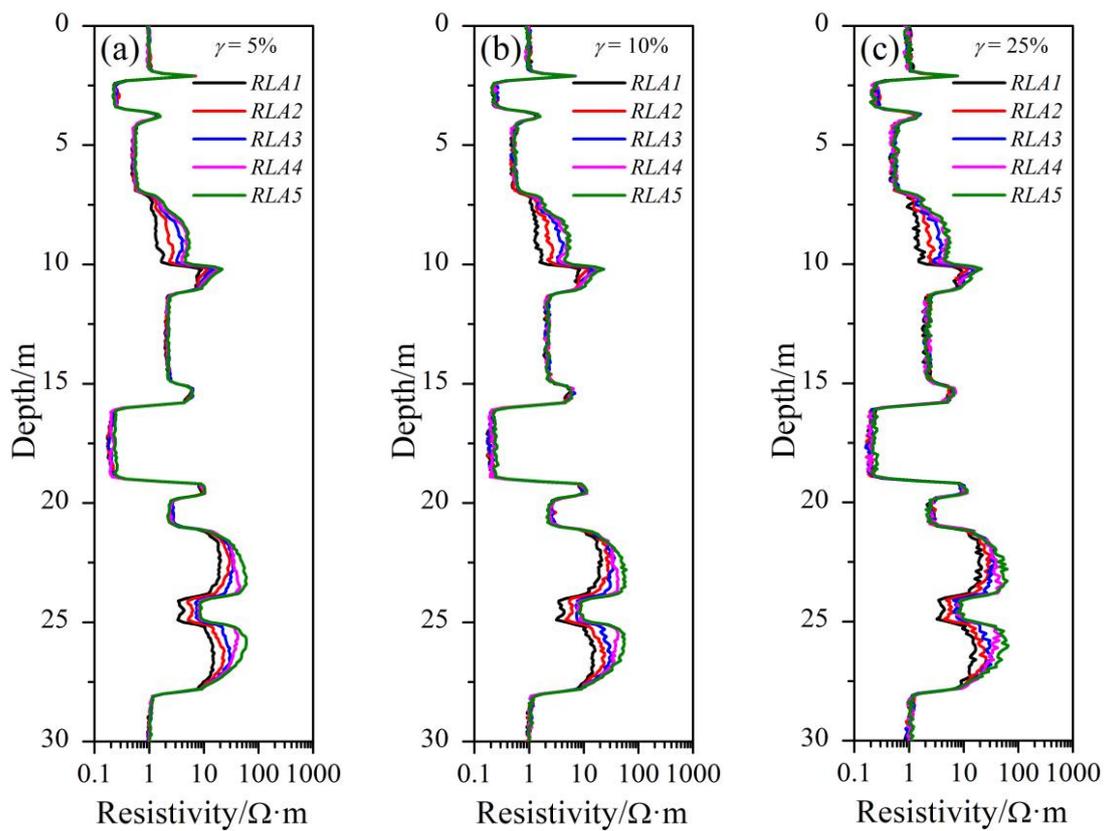

Figure. 12. The array laterolog responses for noise intensities of (a) 5%,(b) 10%,(c) 25%

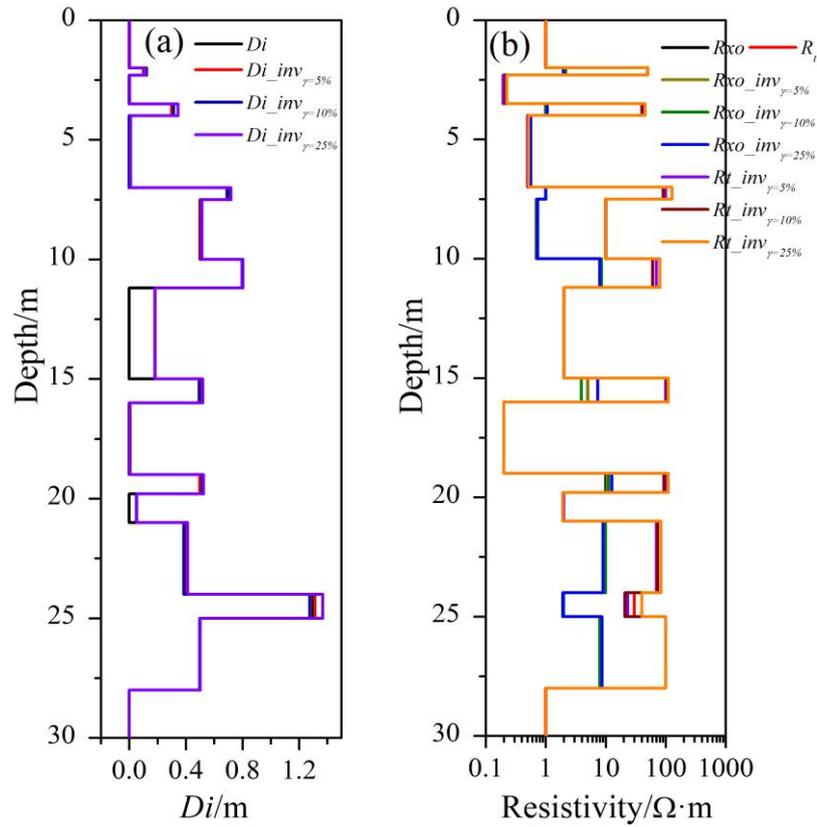

Figure. 13. Inverted results for noise-containing data:(a) Inverted *Di*;(b) Inverted *Rxo* and *Rt*

## 4.3 Uncertainty evaluation of estimated parameters

To evaluate the uncertainty of the inverted results after obtaining the global solution, we introduce 25% pseudorandom noise as an example. Fig. 14 displays the uncertainty of the inverted *Di*, *Rxo* and *Rt*. Notably, the uncertainty of *Di* approaches 1 in some uninvaded formation, e.g., the 1$^{st}$, 9$^{th}$ and 13$^{th}$ beds. In some beds with large invasion depths, such as the fifteenth layer, the uncertainty of *Rt* increases, and the precision of *Rt_inv* decreases. In particular, when the invasion depth approximately 0.4 m, the uncertainties of *Rxo* and *Rt* are relatively close. Overall, the uncertainty of *Rt* is greater than that of *Rxo*.

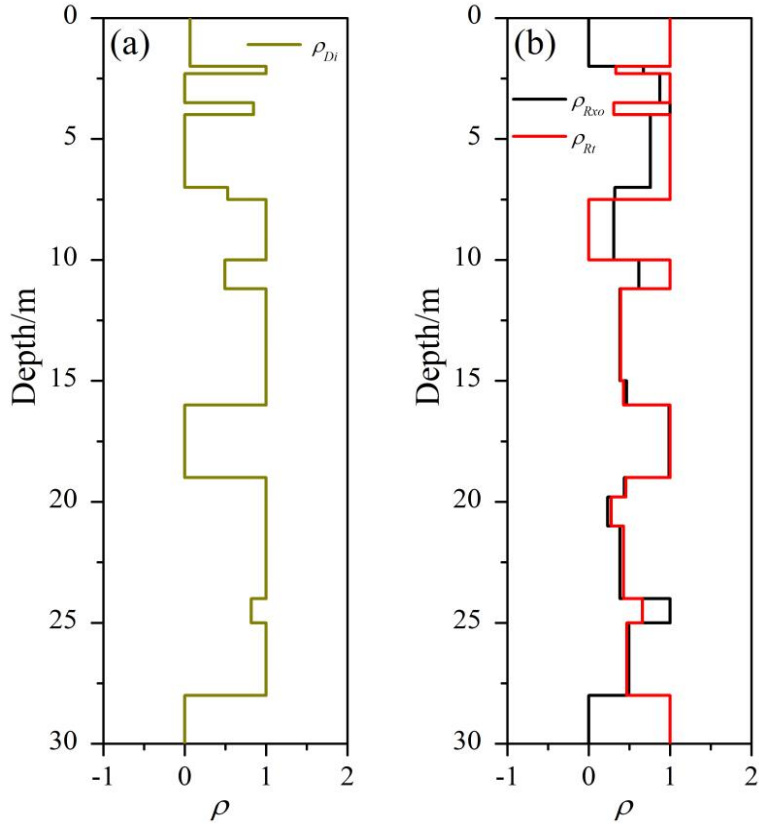

Figure. 14. Uncertainly evaluation of (a) inverted $D_i$;(b) inverted $R_{xo}$ and $R_t$

## 5 Inversion over a three-dimensional anisotropic formation

Because the size of sedimentary particles is irregular and thin interbeds develops, the anisotropic characteristics are obvious in reservoirs. Therefore, in this section, we extend the inversion scheme to 3D anisotropic formations to investigate its efficiency.

*5.1 Responses in deviated anisotropic formation*

Fig. 15 shows the responses of the HRLA in an anisotropic formation with different dipping angles, and the dipping angle ranges from 0º to 90º. The formation is penetrated by a borehole with a diameter of 8 inches, and the borehole is filled with mud with a resistivity of 0.1 Ω.m. The horizontal resistivity $R_h$ is 10 Ω.m. The anisotropic coefficients in Fig. 15a and Fig. 15b are 2 and 5, respectively. The curves with different DOIs become dispersed with increasing anisotropic coefficient. The differences among the five curves reverse and become positive (*RLA1>RLA2>RLA3>RLA4>RLA5*). The reversal angle is located in the range of 40º-65º.

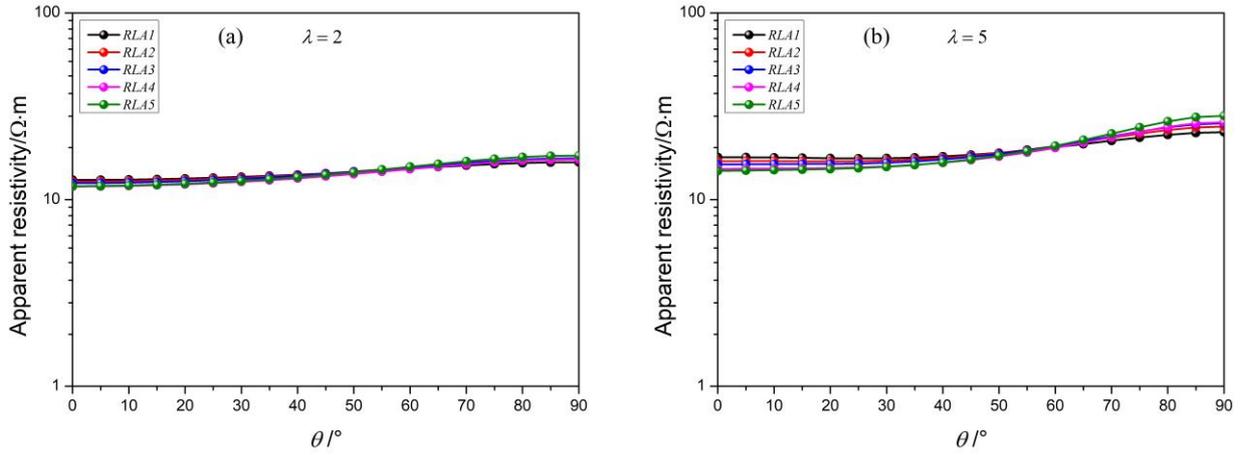

Fig. 15. Influence of anisotropy on the responses of array laterolog:(a) $\lambda=2$;(b) $\lambda=5$

*5.2 ADNN modeling verification for a deviated anisotropic formation*

A three-layer anisotropic model with a dipping angle of $80^0$ is used to verify the suitability of the ADNN approach. In the model, the two surrounding beds are homogeneously isotropic, and the middle bed is homogeneously anisotropic. The thicknesses of the surrounding and middle layers are semi-infinite and 2 m, respectively. The resistivities of two surrounding beds are 2 Ω·m. The horizonal resistivity of the middle layer is 2 Ω·m. Fig. 16 shows a comparison between the data predicted by the ADNN ($RLA\_{ADNN}$) and the data simulated with by 3D-FEM ($RLA\_{3D\text{-}FEM}$). The measurement point is located in middle of the model. The effect of the anisotropic coefficient $\lambda$ on the prediction accuracy is shown in Fig. 16a. The horizontal resistivity is 5 Ω·m, and $\lambda$ ranges from 1 to 2. The maximum relative error between $RLA\_{ADNN}$ and $RLA\_{3D\text{-}FEM}$ is 0.13%. The effect of horizontal resistivity $Rh$ on the prediction accuracy is shown in Fig. 16b. When $\lambda$ is set to 2, the maximum relative error is 0.07%. Therefore, it is feasible to regard ADNN as a surrogate model for 3D-FEM forward simulation.

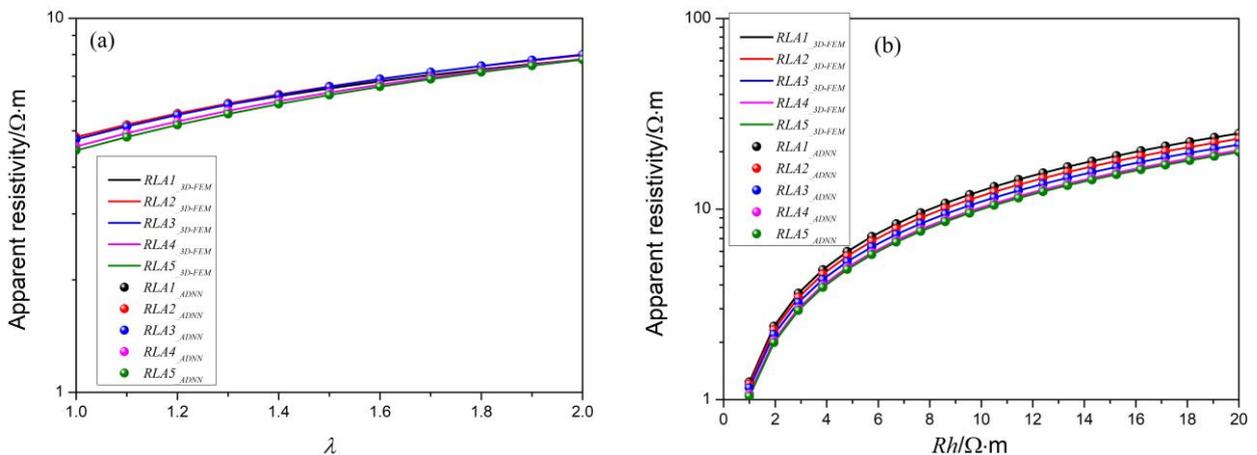

Fig. 16. Comparison of the simulated results of the ADNN and 3D-FEM in anisotropic formations with different (a) $\lambda$ and (b) $Rh$

*5.3 Three-layer inversion experiment*

In this experiment, we apply the inversion scheme to a simple three-layer deviated anisotropic model, and the details of the implementation are the same as those listed in **Sec. 4**. Fig. 17a shows the responses and resistivity distribution of the three-layer model. In the model, the two surrounding beds are homogeneously isotropic, and the middle bed is homogeneously anisotropic. The thicknesses of the surrounding and middle layers are semi-infinite and 3 m, respectively. The resistivities of the two surrounding beds are 2 Ω·m, and the horizontal and vertical resistivities are 20 Ω·m and 100 Ω·m, respectively. The dipping angle is 80°. The inverted results are presented in Fig. 17b. In this case, accurate formation parameters can be obtained after 6 s. From the results, the resistivity profile can be accurately reconstructed, which indicates the robustness of the inversion scheme for 3D formation.

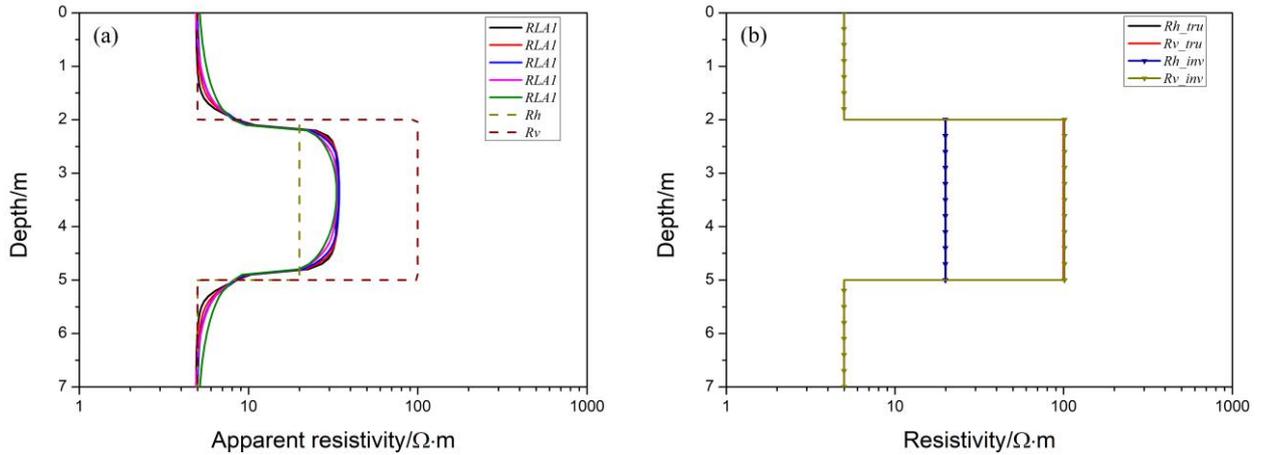

Fig. 17. (a) Model and corresponding array laterolog responses and (b) inverted $Rh$ and $Rt$

## 6 Conclusions

In this paper, a fast hierarchical joint inversion scheme is proposed to extract true formation resistivities from array laterolog measurements in 2D/3D formation with HA/HZ wells. Based on the simulated results and inversion experiments, the following conclusions can be obtained:

(1) A database established with 3D forward simulator is used to develop an ADNN model, which can be used as a surrogate model for 3D-FEM. An adaptive learning rate is introduced into the ADNN architecture to improve the predicted accuracy to 98%, and the model only requires 0.021 s to perform calculations at each logging point.

(2) Compared to the LM algorithm, the AMLM algorithm can optimize anti-noise performance and global convergence in inversion; notably, the proposed method will converge within 15 steps, even if the initial value is 25 times larger than the true value.

(3) A domain decomposition method considering the three-layer model is combined with the AMLM algorithm to improve the inversion accuracy and avoid time consumption in inversion. This approach is nearly 50 times faster than the traditional inversion scheme.

(4) Even at a noise level of 25%, the formation resistivity profile can be accurately reconstructed, and the mean relative error of the inverted $Rt$ values is less than 10%. Moreover, the inversion scheme is robust when applied to anisotropic formations. Although only the HRLA is discussed and processed in this paper, the proposed inversion scheme is also applicable to other borehole measurements.

**Acknowledgements** This work is supported by Natural Science Foundation of China (41674131, 41574118, 41974146, 41904109).